\documentclass[superscriptaddress,twocolumn,showpacs,amsmath,amssymb]{revtex4}
\usepackage[T1]{fontenc}
\usepackage{graphicx}
\usepackage{dcolumn}
\usepackage{bm}
\usepackage{array}
\renewcommand{\vec}[1]{\mbox{\boldmath $#1$}}

\begin{document}

\preprint{}

\title{
Application of random phase approximation to 
vibrational excitations of double-$\Lambda$ hypernuclei}

\author{F. Minato}
\affiliation{
Research group for applied nuclear physics, Japan Atomic Energy Agency,
Tokai 319-1195, Japan}

\author{K. Hagino}
\affiliation{
Department of physics, Tohoku University, Sendai 980-8578, Japan}
\date{\today}

\begin{abstract}
Using 
the Hartree-Fock plus random-phase-approximation (HF+RPA), 
we study the impurity effect of $\Lambda$ hyperon on the collective 
vibrational excitations of double-$\Lambda$ hypernuclei. 
To this end, we employ a Skyrme-type $\Lambda N$ and $\Lambda\Lambda$
interactions for the HF calculations, and the residual interactions for RPA 
derived with the same interactions. 
We find that inclusion of two $\Lambda$ hyperons in $^{16}$O shifts 
the energy of the collective states towards higher energies. 
In particular, the energy of the giant monopole resonance of 
$^{\,\,18}_{\Lambda\Lambda}$O, as well as that of $^{210}_{\Lambda\Lambda}$Pb,  
becomes larger. This implies that 
the effective incompressibility modulus increases 
due to the impurity effect of $\Lambda$ particle, 
if the $\beta$-stability condition is not imposed. 
\end{abstract}
%
\maketitle
\section{Introduction}
Information on the interaction between a $\Lambda$ hyperon and a nucleon 
deepens our 
understanding of baryon-baryon forces and the equation of 
state (EOS) of nuclear matter. In principle, the interaction 
between two particles can be investigated with a measurement of their scattering. 
However, due to the short life-time of $\Lambda$ hyperon, it has yet been difficult 
to perform a direct scattering experiment of nucleon and $\Lambda$ hyperon. 
Therefore, the $\Lambda N$ interaction has been mainly investigated 
by $\gamma$ spectroscopy of single-$\Lambda$ hypernuclei
\cite{Tanida2001,Ukai2006,Tamura2005,Kohri2002,Ukai2004}. 
Such measurements have revealed 
the $\Lambda$-impurity effect, that is, the change of several properties 
of atomic nuclei, such as 
excitation energies and transition probabilities of $\gamma$-ray, due to 
an addition of $\Lambda$ particle. 
Apparently, 
high-resolution $\gamma$-ray measurements 
are vital in investigating $\Lambda$ hypernuclei. 
In addition to the existing experimental data, research projects currently 
planned at the J-PARC facility using new Ge detector arrays 
(Hyperball-J)\cite{Tamura} 
aim at obtaining new data on the low-lying energy level scheme of 
$\Lambda$ hypernuclei in the $sd$ shell region,  
that will lead to further understanding of 
the $\Lambda N$ and $\Lambda\Lambda$ interactions.

Several theoretical calculations have been carried out 
to analyze the relation between low-lying energy 
levels of single-$\Lambda$ hypernuclei and the $\Lambda N$ interaction
\cite{Nemura2002,Millener2008,Umeya2009,Motoba1983,Hiyama1997,
Hiyama2010,Isaka2011,Isaka2011b}.
These calculations 
have not only contributed to identification of energy 
level schemes of single-$\Lambda$ hypernuclei, but have also predicted the 
$\Lambda$-impurity effect on the structure of single-$\Lambda$ hypernuclei,
{\it e.g.}, shrinkage of the radius of $^{7}_{\Lambda}$Li from 
$^6$Li\cite{Motoba1983}, which was subsequently observed experimentally 
\cite{Tanida2001}.

Besides single-$\Lambda$ hypernuclei, double-$\Lambda$ hypernuclei have also been  
studied both experimentally and theoretically. 
Similar to the $\Lambda N$ interaction, information on the $\Lambda\Lambda$ 
interaction can 
be deduced from observation of $\gamma$-rays emitted from double-$\Lambda$ 
hypernuclei.
However, until now double-$\Lambda$ hypernuclei have been produced only in 
an emulsion, and at present 
emitted $\gamma$-rays are difficult to detect experimentally with high 
precision. 
In addition, so far observed double-$\Lambda$ hypernuclei in the emulsion have been 
limited to five cases($^{\,\,\,\,\,6}_{\Lambda\Lambda}$He and 
$^{10-13}_{\quad\,\Lambda\Lambda}$B \cite{Danysz1963,Nakazawa2010}), 
and the experimental data have been scarce. 
Therefore the theoretical approaches make an important tool to assess 
the $\Lambda$ impurity effect on the structure of double-$\Lambda$ hypernuclei 
as well as appropriate selection of a target nucleus for future experiments.
Theoretically, the double-$\Lambda$ hypernuclei have been 
investigated within the frameworks of ab-initio few body model\cite{Nemura2005}, 
shell model\cite{Gal2011} and cluster model\cite{Hiyama2002}.
However, these theoretical approaches demand a huge 
computational power, 
and they may be difficult to apply to heavy hypernuclei. 

In order to study systematically the $\Lambda$-impurity effect, from light 
to heavy nuclei,  
a Hartree-Fock (HF) plus random-phase-approximation (RPA) method provides one of 
the most suitable tools. 
This approach has been applied to study vibrational excitations of normal 
nuclei (without hyperons) 
throughout the nuclear chart, starting with a single energy functional 
applicable in the whole 
range of the nuclear chart. 
In particular, the RPA has been successfully applied to descriptions of 
giant resonances of atomic nuclei. 
See Refs.\cite{AGL77,AG80,LA95,MDMC08} for earlier applications of 
Tamm-Dancoff approximation to (K$^-$,$\pi^-$) and ($\pi^+$,K$^+$) reactions, 
and of an RPA-like model 
to single-particle spectra of single-$\Lambda$ hypernuclei. 
So far, the mean field approach has been extended 
to $\Lambda$ hypernuclei in order 
to study the ground state properties
\cite{Rayet,Yamamoto1988,Lanskoy1998,Tretyakova1999,Schulze1995,Vretenar1998}, 
the potential energy surface in the deformation plane 
\cite{Myaing2008,Myaing2011,LZZ11} and 
fission barrier heights \cite{Minato}.
Concerning excited states,
the low-lying excited states of $^{25}_{\Lambda}$Mg
have recently been calculated with a 5-dimensional (5D) collective 
Bohr Hamiltonian on the basis of 
the Skyrme-Hartree-Fock method \cite{Yao2011} (see also Ref.\cite{Isaka2011} 
for a recent application 
of anti-symmetrized molecular dynamics to the $^{25}_{\Lambda}$Mg hypernucleus).
Although the Bohr Hamiltonian approach can handle a large amplitude 
collective motion, it is much easier 
to employ the RPA approach to describe collective vibrations with 
several multipolarities, including 
giant resonances as well.

In this paper, we 
extend the Skyrme-HF plus RPA (SHF+RPA) scheme to hypernuclei.  
Skyme-type $\Lambda$N and $\Lambda\Lambda$ interactions, similarly to 
the Skyrme nucleon-nucleon interaction, 
are used in this work. 
The residual interactions for RPA are derived 
self-consistently from the second derivative of the 
energy functional with respect to densities.
In this study, we shall focus on the double-$\Lambda$ hypernuclei 
rather than single-$\Lambda$ hypernuclei, partly because 
the description is much simpler due to the time-reversal symmetry. 
The $\Lambda$-impurity effect is expected to be stronger in 
double-$\Lambda$ hypernuclei, and such calculations will provide the upper limit of 
the impurity effect for single-$\Lambda$ hypernuclei.

The paper is organized as follows.
In Sec. II, we describe the formalism of the SHF+RPA for hypernuclei. 
In Sec. III, we apply the SHF+RPA method to 
$^{\,\,18}_{\Lambda\Lambda}$O hypernuclei and present 
the results for the strength distributions 
and the transition densities for 
the isoscalar giant monopole resonance (IS $0^+$), the electric dipole (E1), 
quadrupole (E2) and octupole (E3) transitions. 
We also calculate the isoscalar giant monopole resonance 
of $^{210}_{\Lambda\Lambda}$Pb hypernucleus 
and discuss the nuclear incompressibility in the presence of $\Lambda$ hyperon. 
We give a summary of the paper in Sec. IV.

\section{RPA FOR HYPERNUCLEI}
In order to describe the ground state and excited states of double-$\Lambda$ 
hypernuclei, 
we adopt the Skyrme-type zero-range force
for the $\Lambda N$ and $\Lambda\Lambda$ interactions.
The $\Lambda N$ and 3-body $\Lambda NN$ interactions of this type were 
first introduced by Rayet as\cite{Rayet},
\begin{equation}
\begin{split}
&v_{\Lambda N}(\vec{r}_\Lambda-\vec{r}_N)
=
t_0^\Lambda(1+x_0^\Lambda P_\sigma)\delta(\vec{r}_{\Lambda}-\vec{r}_N)\\
&+\frac{1}{2}t_1^\Lambda\left(\vec{k'}^2\delta(\vec{r}_{\Lambda}-\vec{r}_N)
+\delta(\vec{r}_{\Lambda}-\vec{r}_N)\vec{k}^2\right)\\
&+t_2^\Lambda\vec{k'}\delta(\vec{r}_\Lambda-\vec{r}_N)\cdot\vec{k}
+iW_0^\Lambda\vec{k'}\delta(\vec{r}_\Lambda-\vec{r}_N)\cdot(\sigma\times\vec{k}),
\end{split}
\label{LN}
\end{equation}
and 
\begin{equation}
v_{\Lambda NN}(\vec{r}_\Lambda,\vec{r}_{N_1},\vec{r}_{N_2})=
t_3^\Lambda\delta(\vec{r}_\Lambda-\vec{r}_{N_1})
\delta(\vec{r}_{\Lambda}-\vec{r}_{N_2}),
\label{LNN}
\end{equation}
respectively. In a similar way, Lanskoy introduced the $\Lambda\Lambda$ 
interaction as \cite{Lanskoy1998}, 
\begin{equation}
\begin{split}
v_{\Lambda\Lambda}&(\vec{r}_{\Lambda_1}-\vec{r}_{\Lambda_2})
=
\lambda_0\delta(\vec{r}_{\Lambda_1}-\vec{r}_{\Lambda_2})\\
&+\frac{1}{2}
\lambda_1\left(\vec{k'}^2\delta(\vec{r}_{\Lambda_1}-\vec{r}_{\Lambda_2})
+\delta(\vec{r}_{\Lambda_1}-\vec{r}_{\Lambda_2})\vec{k}^2\right)\\
&+\lambda_2\vec{k'}\delta(\vec{r}_{\Lambda_1}-\vec{r}_{\Lambda_2})\cdot\vec{k}\\
&+\lambda_3\,
\left[\rho_N\left(\frac{\vec{r}_{\Lambda_1}+\vec{r}_{\Lambda_2}}{2}\right)
\right]^{\alpha_\Lambda}\,\delta(\vec{r}_{\Lambda_1}-\vec{r}_{\Lambda_2}).
\end{split}
\label{LL}
\end{equation}
The operator $\vec{k'}=-(\vec{\nabla}_1-\vec{\nabla}_2)/2i$ acts on the left
hand side while $\vec{k}=(\vec{\nabla}_1-\vec{\nabla}_2)/2i$ acts on 
the right hand side.
$\rho_N(\vec{r})$ is the density distribution for the nucleons. 
The last term in Eq.\eqref{LL} corresponds to the three-body $\Lambda\Lambda N$
interaction, 
which originates mainly from the $\Lambda\Lambda-\Xi N$ 
coupling \cite{Hiyama2004}.

Together with the Skyrme $NN$ interaction\cite{Vautherin72}, 
the total energy $E_{\rm tot}$ in the Hartree-Fock approximation 
is given by
\begin{equation}
E_{\rm tot}=E_N+E_\Lambda,
\label{energy}
\end{equation}
where
\begin{equation}
E_N=\int H_N(\vec{r})\,d\vec{r},
\label{energy2}
\end{equation}
is the energy for the core nucleus without $\Lambda$ hyperons while 
\begin{equation}
E_\Lambda=\int [H_{N\Lambda}(\vec{r})+H_{\Lambda\Lambda}(\vec{r})]\,d\vec{r},
\label{energy3}
\end{equation}
is due to the 
$\Lambda N$ 
and $\Lambda\Lambda$ interactions (see Appendix A).
The kinetic energy density for $\Lambda$ particles is included in 
the energy density $H_{N\Lambda}(\vec{r})$. 

The SHF equations are obtained by taking variation of the total energy 
$E_{\rm tot}$ with respect to the densities 
for neutrons, protons and $\Lambda$ hyperons. 
These are given as 
\begin{equation}
\left(-\vec{\nabla}\cdot 
\frac{\hbar^2}{2m_b^*(\vec{r})}\vec{\nabla}+U_{bN}(\vec{r})
+U_{b\Lambda}(\vec{r})\right)\phi_b(\vec{r})=\epsilon_b\phi_b(\vec{r}),
\label{shfeq1}
\end{equation}
where the index $b$ refers to proton, neutron or $\Lambda$, 
and $\epsilon_b$ is the single-particle energy. 
The explicit forms for the 
mean-field potentials $U_{bN}(\vec{r})$ and $U_{b\Lambda}(\vec{r})$, 
and the effective mass $m_b^*(\vec{r})$ are given in Appendix A. 

After we construct the ground state in the Hartree-Fock approximation, 
we describe excited states with RPA as a linear superposition of 
1 particle-1 hole (1p1h) configurations. That is, the excitation operator 
$Q_k^\dagger$ 
for the $k$-th RPA phonon is assumed to be, 
\begin{equation}
Q_k^\dagger = \sum_{p,h\in n,p,\Lambda}\left(
X_{ph}^{(k)}a_p^\dagger a_h - 
Y_{ph}^{(k)}a_h^\dagger a_p\right), 
\end{equation}
where $X_{ph}^{(k)}$ and $Y_{ph}^{(k)}$ are
the forward and backward amplitudes, respectively. 
$a_p^\dagger$ and $a_h^\dagger$ are the creation operators for a particle state 
$p$ and for a hole state $h$, respectively. 
The excitation energy $E_k$ is obtained by diagonalizing 
the $2\nu$-dimensional RPA equation,
\begin{equation}
\left(\begin{tabular}{cc}
$A$ & $B$\\
$-B^*$ & $-A^*$
\end{tabular}\right)
\left(\begin{tabular}{c}
$X^{(k)}$\\$Y^{(k)}$
\end{tabular}\right)
=E_k
\left(\begin{tabular}{c}
$X^{(k)}$\\$Y^{(k)}$
\end{tabular}\right),
\label{rpa}
\end{equation}
where $\nu$ is the number of 1p1h configurations. 
Here, 
$A$ and $B$ are RPA matrices given by, 
\begin{equation}
\begin{split}
A_{ph,p'h'}&=(\epsilon_p-\epsilon_h)\delta_{pp'}\delta_{hh'}+v_{ph'hp'}\\
B_{ph,p'h'}&=v_{pp'hh'}, 
\end{split}
\label{abmatrix}
\end{equation}
where $v$ is the residual interaction derived from the energy functional, 
$E_{\rm tot}$. 
The formalism is almost the same as the standard RPA found {\it e.g.,}  in 
Refs.\cite{RingSchuck,Rowe}, but the particle-hole configurations 
run over not only protons and neutron but also $\Lambda$ hyperons.
The interaction matrix elements $v_{ph'hp'}$ and $v_{pp'hh'}$ include
the $\Lambda N$ and $\Lambda\Lambda$ interactions 
as well as the $NN$ interaction (see Appendix B).

The external fields for electric multipole excitations with multipolarities 
$L\neq$ 0 and 1 are defined as
\begin{equation}
\hat{F}_{EL}=e\sum_{i\in p} r_i^LY_{LM}(\hat{r_i}), 
\end{equation}
while that for the ``isoscalar'' monopole transition is
\begin{equation}
\hat{F}_{0^+}=\sum_{i\in p,n,\Lambda} r_i^2.
\end{equation}
For the electric dipole response, 
we take into account the center of mass motion of the whole hypernucleus 
and use the 
operator 
\begin{equation}
\begin{split}
\hat{F}_{E1}
&=e\sum_{i\in p}(r_iY_{1M}(\hat{r}_i)-RY_{1M}(\vec{R})), \\
&=e\frac{Nm_N+N_\Lambda m_\Lambda}{M}\sum_{i\in p}r_iY_{1M}(\hat{r}_i)\\
&-e\frac{Z}{M}\left(m_N\sum_{i\in n}r_iY_{10}(\hat{r}_i)
+m_\Lambda\sum_{i\in \Lambda} r_i Y_{10}(\hat{r}_i)\right),
\end{split}
\end{equation}
where 
\begin{equation}
\vec{R}=\frac{1}{M}\,\left(m_N\sum_{i\in n,p} \vec{r}_i + m_\Lambda\sum_{i\in \Lambda} 
\vec{r}_i\right), 
\end{equation}
is the center of mass of the hypernucleus,  
and 
$M \equiv m_N(Z+N)+m_\Lambda N_\Lambda$ is the total mass, 
$m_N=(m_p+m_n)/2=938.92$ MeV/$c^2$ and $m_\Lambda=1115.68$ MeV/$c^2$
being the mass of nucleon and $\Lambda$ hyperon, respectively.
$N$, $Z$ and $N_\Lambda$ are the number of neutron, proton and
$\Lambda$ hyperon, respectively.

\section{RESULTS}
\subsection{single-particle level of $^{\,\,18}_{\Lambda\Lambda}$O}
\begin{figure}
\includegraphics[width=0.7\linewidth]{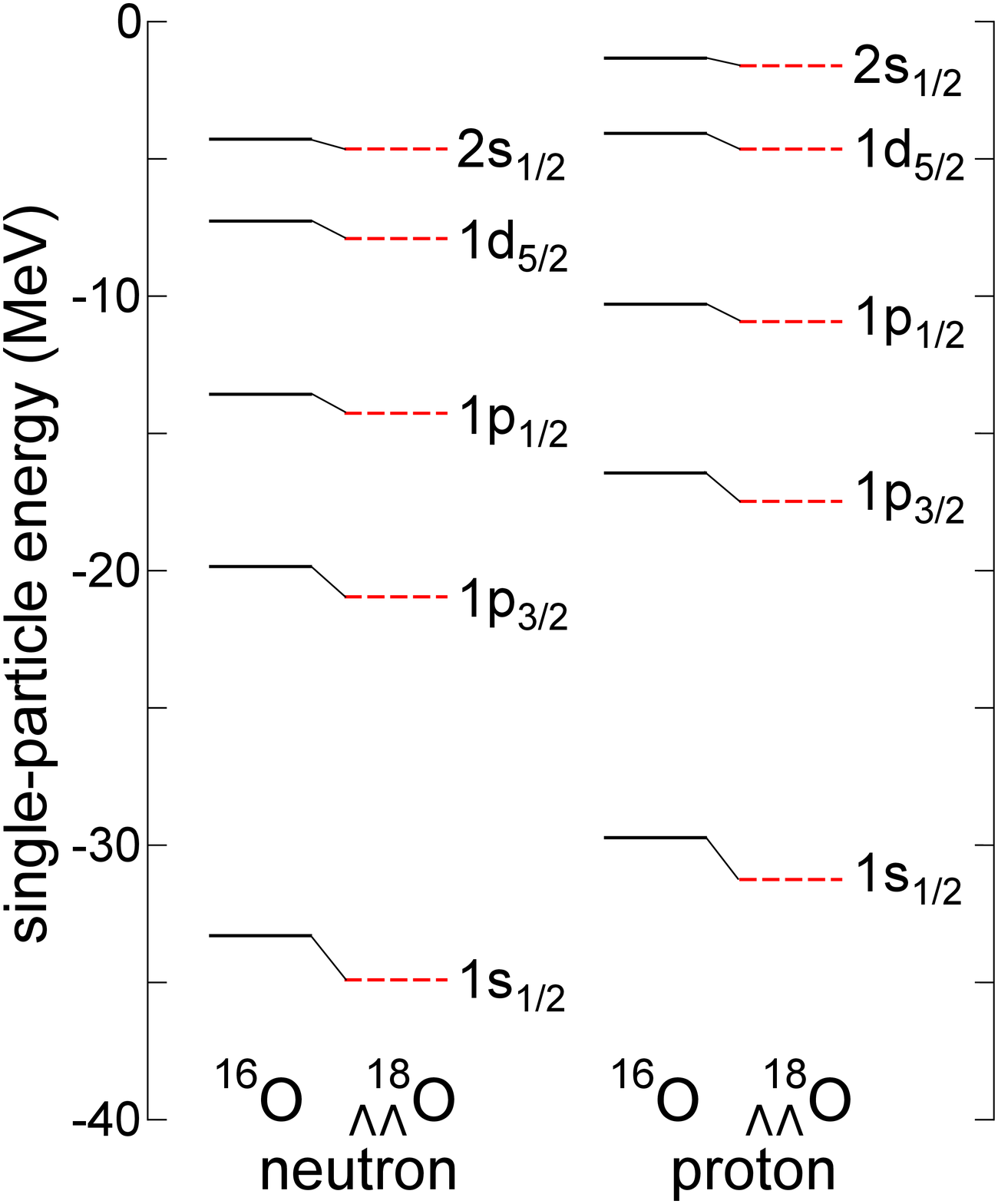}
\caption{(Color online) Neutron and proton single-particle levels 
of $^{16}$O (the solid lines) and $^{\,\,18}_{\Lambda\Lambda}$O (the dashed lines) 
obtained with the Skyrme-Hartree-Fock method.}
\label{single-particle}
\end{figure}

We now numerically solve the RPA equation and discuss the collective 
excitations of double-$\Lambda$ hypernuclei. 
Before we show the results for multipole vibrations,
we first discuss the single particle levels of double-$\Lambda$ hypernucleus
$^{\,\,18}_{\Lambda\Lambda}$O and $^{16}$O,
which will help to understand the 
$\Lambda$-impurity effect on the giant resonances.
To this end, 
we assume spherical symmetry, and 
solve the SHF equation in the coordinate space with a grid size of 
$dr$=0.1 fm. 
We use the SkM$^*$ parameter set for $NN$ interaction \cite{SkM*},  
while the No.5 parameter set in Ref.\cite{Yamamoto1988} for the $\Lambda N$
interaction, whose parameters were determined by fitting the Hartree-Fock 
calculations to 
the experimental binding energies of 
single-$\Lambda$ hypernuclei\cite{Yamamoto1988}. 
For the $\Lambda\Lambda$ interaction, we use the S$\Lambda\Lambda1$ parameter
set evaluated by Lanskoy\cite{Lanskoy1998}.
This parameter set was obtained by fitting 
to the 
$\Lambda\Lambda$ bond energy \cite{Lanskoy1998},
$\Delta B_{\Lambda\Lambda}=B_{\Lambda\Lambda}-2B_{\Lambda}$, 
where $B_{\Lambda}$ is the one-$\Lambda$ hyperon separation energy
from a $^{\rm{A}+1}_{\quad\Lambda}$Z hypernucleus
and $B_{\Lambda\Lambda}$ is the two-$\Lambda$ hyperon separation energy 
of $^{\rm{A}+2}_{\,\,\,\Lambda\Lambda}$Z.
As we will show in the next subsection, the dependence of giant resonances 
on a choice of parameter sets for the $\Lambda N$ and the 
$\Lambda\Lambda$ interactions is weak,
and any significant change in the results is 
not obtained even if we use different parameter sets for the interactions.

Figure \ref{single-particle} shows the neutron and proton single-particle 
levels of the $^{16}$O (the solid lines) and the $^{\,\,18}_{\Lambda\Lambda}$O 
(the dashed lines).
The single-particle energies of the $^{\,\,18}_{\Lambda\Lambda}$O hypernucleus 
are smaller 
than those of $^{16}$O, since 
the depths of the central part of the mean-field potentials are deepened 
both for protons and neutrons 
due to the attractive 
$\Lambda N$ interaction, 
as shown in Fig.2 of Ref.\cite{Tretyakova1999}. 
The lowest level ($1s_{1/2}$) is the most sensitive to the addition of 
$\Lambda$ particles, 
for which the single particle levels are shifted by 
$\Delta\epsilon_n=-1.4$ MeV for 
neutron and $\Delta\epsilon_p=-1.3$ MeV for proton.
This $\Lambda$-impurity effect becomes weaker 
as the energy of a single-particle state increases. 
If the continuum spectra are discretized within a large box, 
the difference of single-particle energies 
between the $^{16}$O and the $^{\,\,18}_{\Lambda\Lambda}$O nuclei 
is much smaller 
as compared to the bound levels. 
Consequently, 
in the independent-particle approximation, 
the excitation energies increase relatively in the 
double-$\Lambda$ hypernucleus as compared to those of the normal nucleus. 
In the next subsections, we will see that this is the case even in the 
presence of the residual particle-hole interaction.  

\subsection{Low-lying excitations}

We next solve the RPA equation in order to discuss collective excitations 
of the $^{\,\,18}_{\Lambda\Lambda}$O hypernucleus. 
To this end, 
we discretize the single-particle continuum states with the box boundary condition
with the box size of $16.0$ fm. We take into account the continuum states 
up to $\epsilon$=30 MeV, and 
consider the 1p1h configurations whose 
unperturbed energy, $\epsilon_p-\epsilon_h$, is smaller than 60 MeV. 
For the residual interactions, 
we neglect the Coulomb and the spin-orbit terms for simplicity, although 
we include all the other terms self-consistently. 
Therefore, our RPA calculations are not fully self-consistent, and 
the spurious translational motion appears at a finite excitation energy. 
In order to recover effectively the self-consistency, 
we introduce a scaling factor $f$ to the residual 
interaction $v_{res}$ so as to produce the spurious translational mode at 
zero energy.

Table \ref{1stE2E3} shows the results of such RPA calculations for the lowest 
quadrupole and the octupole states of 
$^{16}$O and $^{\,\,18}_{\Lambda\Lambda}$O. 
For both the 2$_1^+$ and 3$_1^-$ states, the impurity effects of 
$\Lambda$ particles slightly reduces the collectivity, that is, 
the excitation energies are increased while the electromagnetic transition 
probabilities are decreased by 26-28 \%. 
The increase of the excitation energies is consistent with the increase of unperturbed particle-hole 
energies discussed in the previous subsection. 

\begin{table}
\caption{The excitation energies and the electromagnetic transition probabilities, 
$B(E2)$ and $B(E3)$, for the first 2$^+$ and 3$^-$ states of 
$^{16}$O and $^{\,\,18}_{\Lambda\Lambda}$O nuclei obtained with the 
Skyrme-HF+RPA method.}
\begin{tabular}{c|cc|cc}
\hline\hline
     & \multicolumn{2}{c|}{2$_1^+$ state} & \multicolumn{2}{c}{3$_1^-$ state}\\
\hline
nucleus & $E$ (MeV) & $B(E2)$ (e$^2$fm$^4$) & $E$ (MeV) & $B(E3)$ (e$^2$fm$^6$)\\
\hline
$^{16}$O                   & $13.1$ & $0.726$ & $6.06$ & $91.1$ \\
$^{\,\,18}_{\Lambda\Lambda}$O & $13.8$ & $0.529$ & $6.32$ & $67.7$ \\
\hline\hline
\end{tabular}
\label{1stE2E3}
\end{table}

\subsection{Giant resonances}
\begin{figure}
\includegraphics[width=0.70\linewidth]{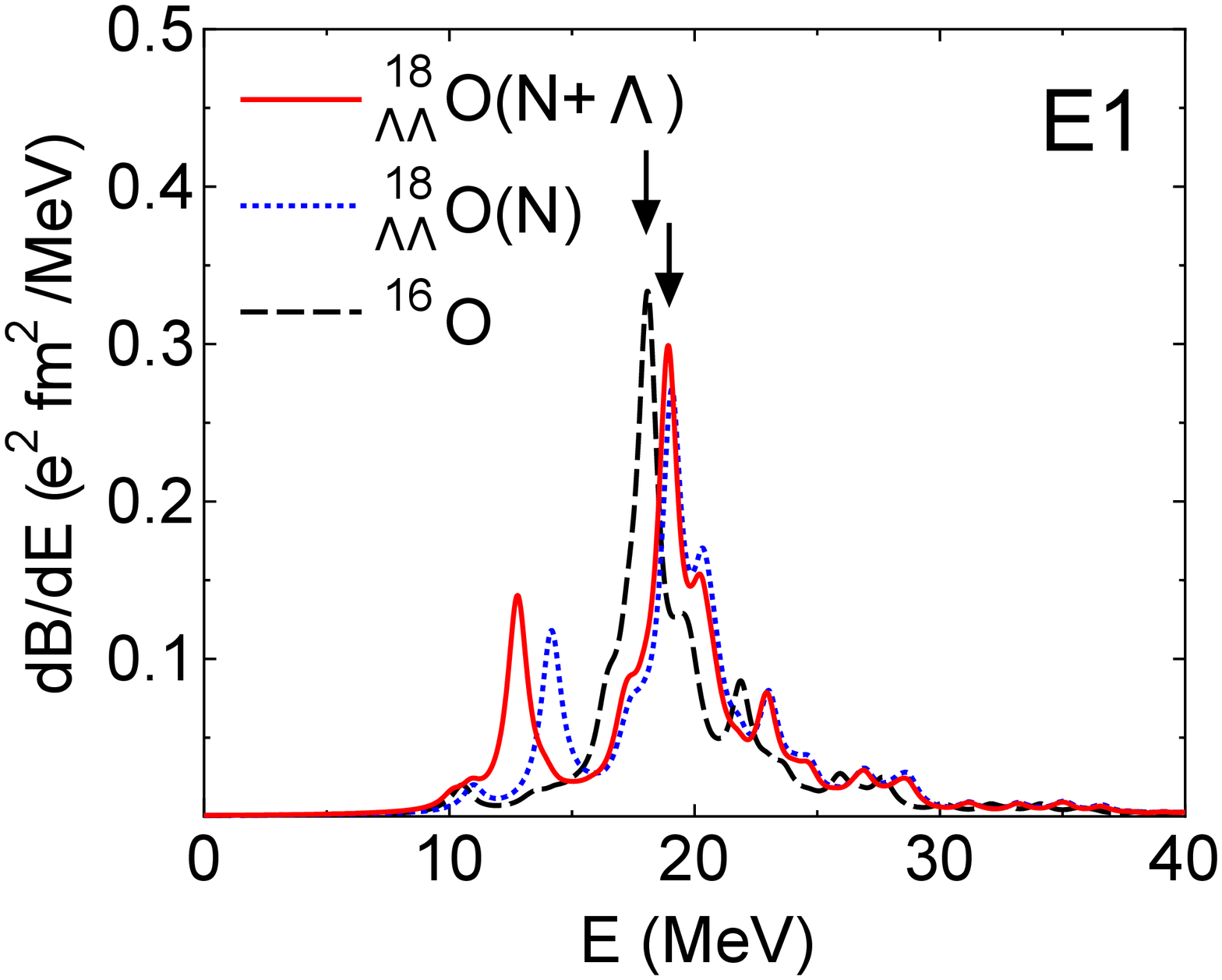}
\includegraphics[width=0.70\linewidth]{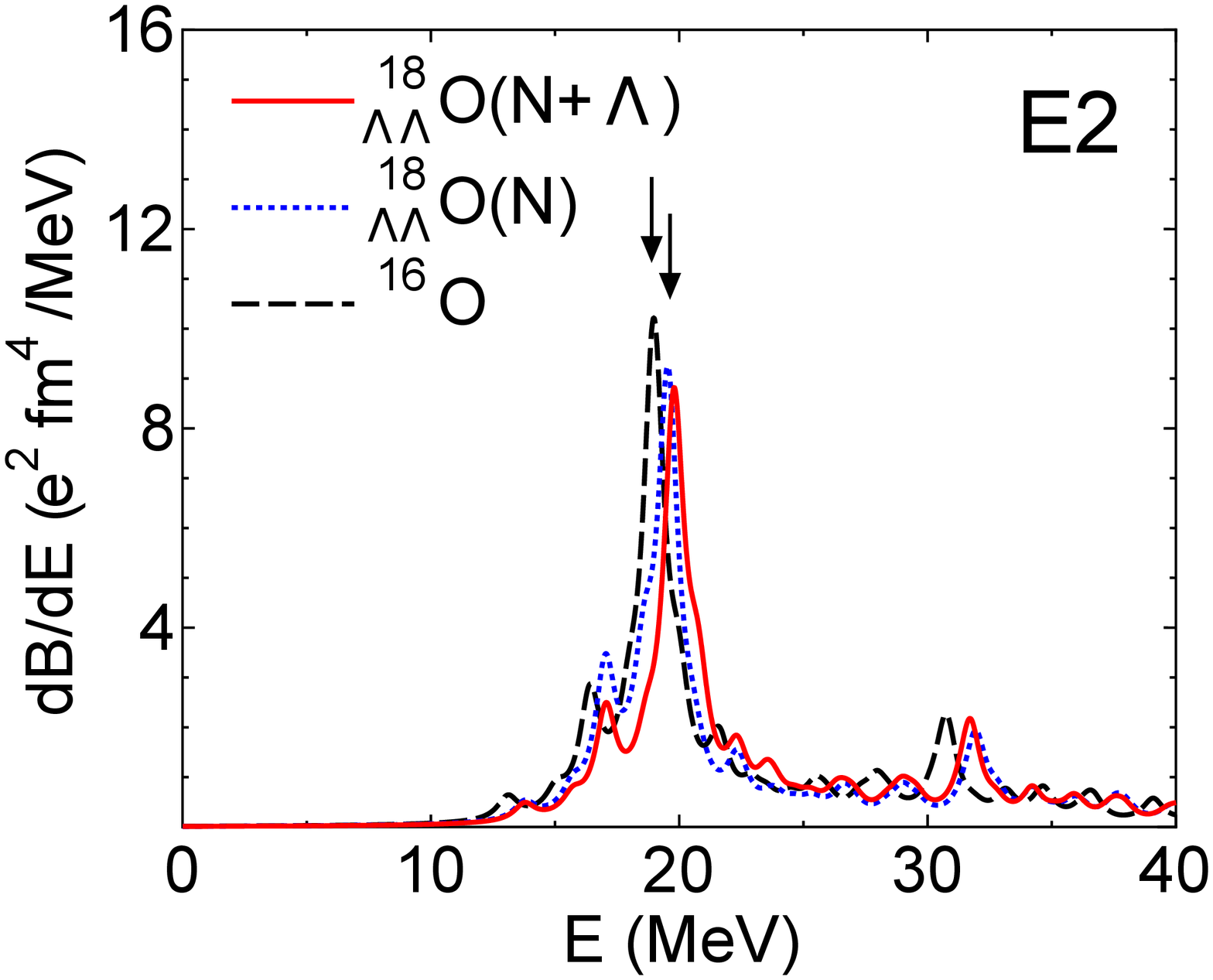}
\includegraphics[width=0.70\linewidth]{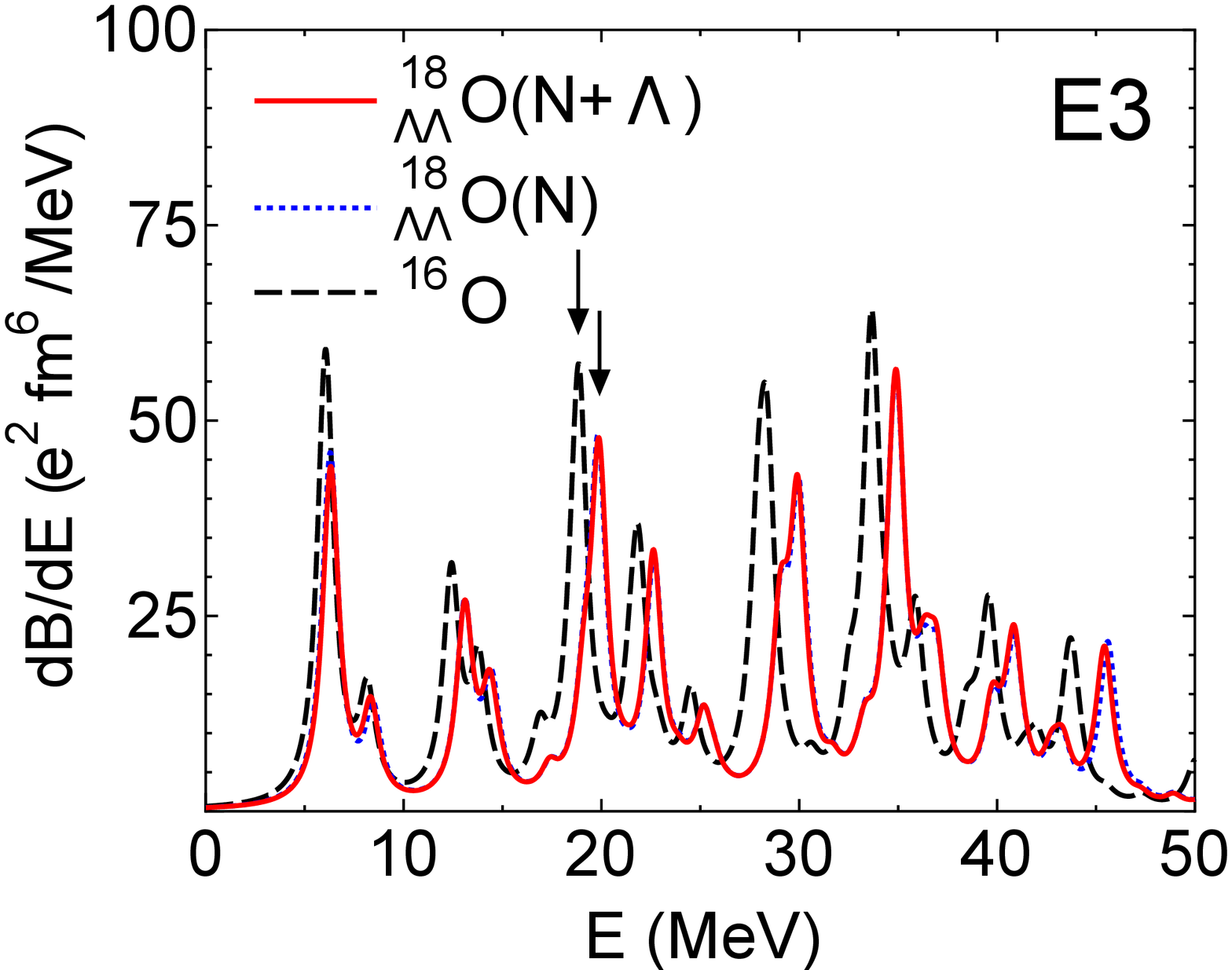}
\caption{(Color online) The strength distributions for the electric dipole 
(E1, the top panel), the electric quadrupole (E2, the middle panel) 
and the electric octupole (E3, the bottom panel) excitations 
of the $^{16}$O nucleus (the dashed lines) and of the 
double-$\Lambda$ hypernucleus $^{\,\,18}_{\Lambda\Lambda}$O 
(the solid and the dotted lines).
The solid lines are obtained by 
including 
the residual $NN$, $\Lambda N$ and $\Lambda\Lambda$ interactions, 
while the dotted lines are obtained by including only 
the residual $NN$ interactions. 
The strength distributions are weighted by the Lorentzian 
function with a width of $1.0$ MeV.
For the peaks indicated by the arrows, the transition densities are 
shown in Fig.\ref{drho_O16}.}
\label{multi}
\end{figure}

The RPA method has been successfully applied to giant resonances of the normal 
nuclei. We therefore apply it in this subsection to the giant resonances of 
the double-$\Lambda$ hypernucleus $^{\,\,18}_{\Lambda\Lambda}$O, although 
they may not be easy to access experimentally at present. 
The top, the middle, and the bottom panels of Fig. \ref{multi} show 
the strength distributions 
for the electric dipole (E1), quadrupole (E2) and octupole (E3) 
excitations, respectively, 
weighted by the Lorentzian function with a width of $1.0$ MeV.
The solid and the dashed lines denote the results for the 
$^{\,\,18}_{\Lambda\Lambda}$O and $^{16}$O nuclei, respectively. 
In order to assess the role of 
$\Lambda$ hyperon, we also show the results for 
the $^{\,\,18}_{\Lambda\Lambda}$O 
hypernuclei in which 
the $\Lambda N$ and $\Lambda\Lambda$ interactions 
are taken into account only in the ground state, that is, 
the results obtained by switching off the 
residual $\Lambda N$ and $\Lambda\Lambda$ interactions (the dotted lines). 
The figure indicates that 
the addition of the $\Lambda$ hyperons shifts the peaks of the strength 
functions towards high energy for all the modes of excitations. 
This is similar to the results for the low-lying modes of excitations discussed in 
the previous subsection, and is again caused mainly by the change of the 
single-particle energies. 
On the other hand, the difference between 
the solid and the dotted lines 
is relatively small, except for the low-lying 
dipole state at $E=12.8$ MeV.
We have confirmed that the strength functions remain almost the same, 
including the low-lying dipole state, even if other parameter sets for 
the $\Lambda N$ and $\Lambda\Lambda$ interactions are employed. 
This suggests that 
the main effect of $\Lambda$ particles on the 
collective vibrational excitations 
is indeed attributed to the change in the single-particle energies, 
rather than the residual $\Lambda N$ and 
$\Lambda\Lambda$ interactions, although 
the low-lying dipole state may require a separate analysis (see Fig. 4 below).

In order to see the $\Lambda$-impurity effect quantitatively,
we show in Table \ref{centroid} 
the centroid energy defined as $\overline{E}=m_1/m_0$, 
where $m_k$ is $k$-th energy-weighted sum-rule, 
\begin{equation}
m_k=\sum_{\nu}(E_i)^k \Big|\langle i|F|0\rangle\Big|^2, 
\label{ewsr}
\end{equation}
for the unperturbed (HF) and the perturbed (RPA) strength functions. 
We also list the difference of the centroid energy, 
$\delta\overline{E}$, 
between 
the $^{\,\,18}_{\Lambda\Lambda}$O and the $^{16}$O nuclei. 
The values in the parentheses for the RPA E1 response 
are the results obtained by excluding 
the contribution of the low-lying 
dipole peak at $E=12.8$ MeV.
As is expected, the centroid energies  
$\overline{E}$ for the HF calculations 
shift to higher energies when the $\Lambda$ hyperons are added to 
$^{16}$O.
For the E2 and E3 responses, this remain the same even if 
the residual interactions are taken into account in RPA. 
For the E1 response, the energy shift is negative, 
but it turs to positive if the low-lying peak is excluded. 
As we will show below, this low-lying peak 
corresponds to the dipole motion of 
a $\Lambda$ hyperon against the core nucleus. 
We thus conclude that 
the $\Lambda$ hyperons generally increases the centroid energy for collective 
motions of the core nucleus, not only for the quadrupole and the octupole 
states but also for the dipole states. 

\begin{table}
\caption{The centroid energy $\overline{E}$ for the E1, E2 and E3
modes of excitations for $^{16}$O and $^{\,\,18}_{\Lambda\Lambda}$O nuclei. 
Those are given in units of MeV, and the results of 
both 
the unperturbed (HF) and the perturbed (RPA) 
calculations are shown. 
$\delta \overline{E}$ denotes the difference of the centroid 
energies between $^{\,\,18}_{\Lambda\Lambda}$O and $^{16}$O. 
The values in the parentheses for the E1 mode 
are the results obtained by excluding the 
contribution of the low-lying peak at $E=12.8$ MeV.}
\begin{tabular}{r|r|ccc}
\hline 
\hline
&& E1 & E2 & E3 \\
\hline
(HF)&$^{16}$O                   & $13.76$ & $25.57$ & $26.53$ \\
    &$^{\,\,18}_{\Lambda\Lambda}$O & $14.34$ & $26.63$ & $27.74$ \\
\cline{2-5}
    &$\delta \overline{E}$ & $+0.58$ & $+1.06$ & $+1.21$ \\
\hline
(RPA)&$^{16}$O                  & $19.92$ & $19.55$ & $22.32$ \\
     &$^{\,\,18}_{\Lambda\Lambda}$O& $19.68$ ($20.95$) & $20.09$ & $24.05$\\
\cline{2-5}
     &$\delta \overline{E}$ & $-0.24$ ($+1.03$) & $+0.54$ & $+1.73$ \\
\hline
\hline
\end{tabular}
\label{centroid}
\end{table}

\begin{figure*}
\includegraphics[width=0.300\linewidth]{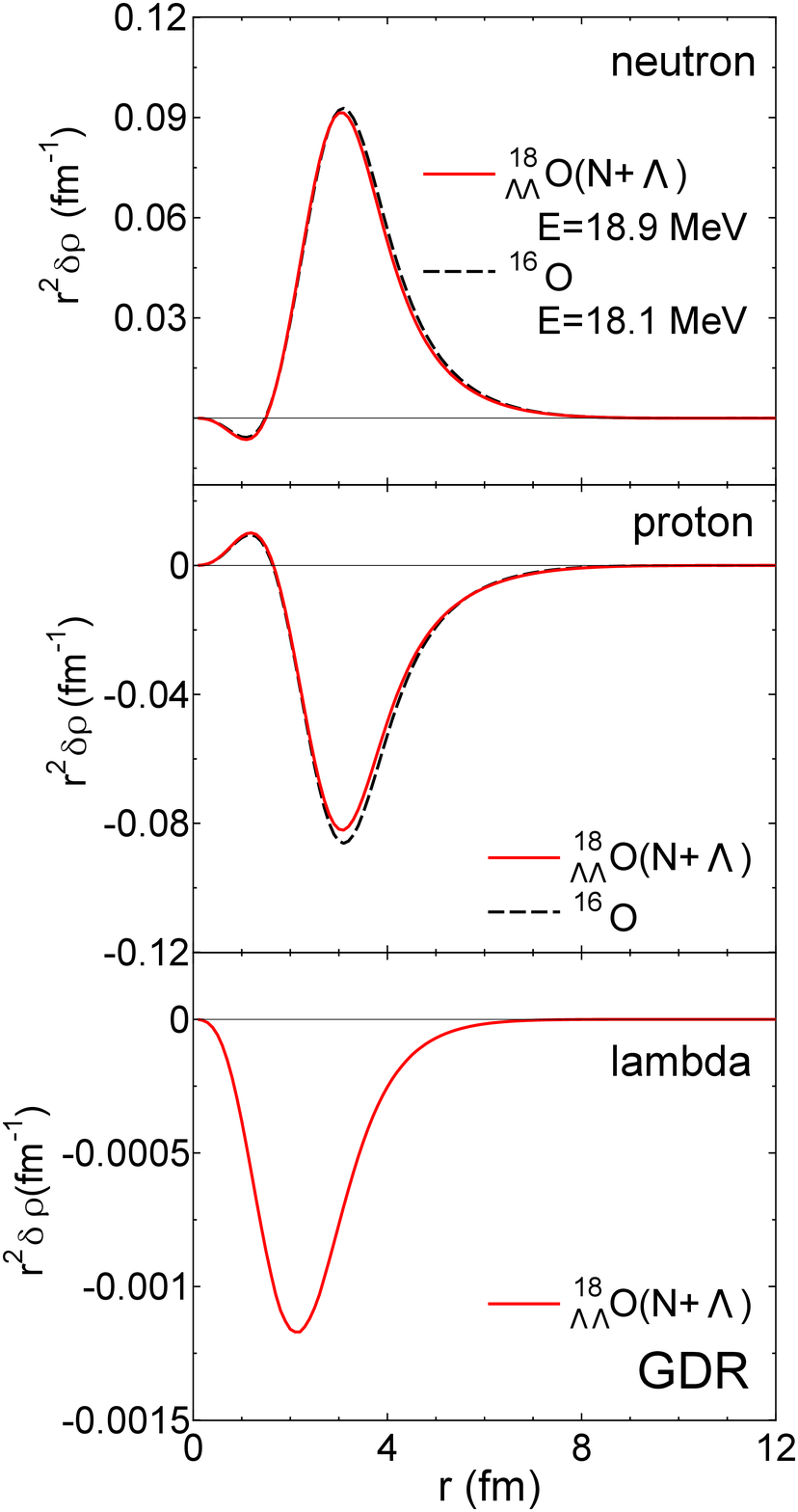}
\includegraphics[width=0.286\linewidth]{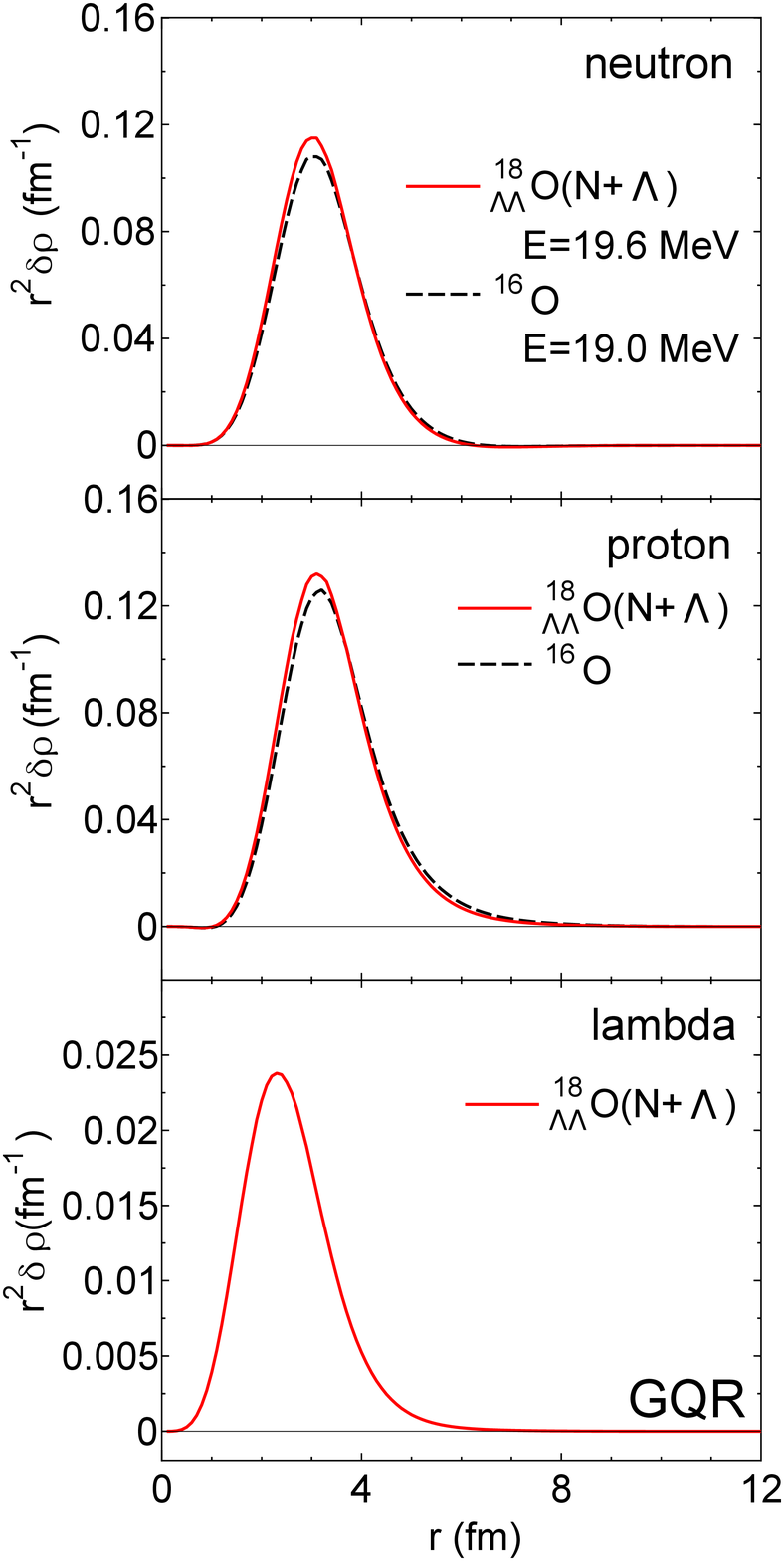}
\includegraphics[width=0.292\linewidth]{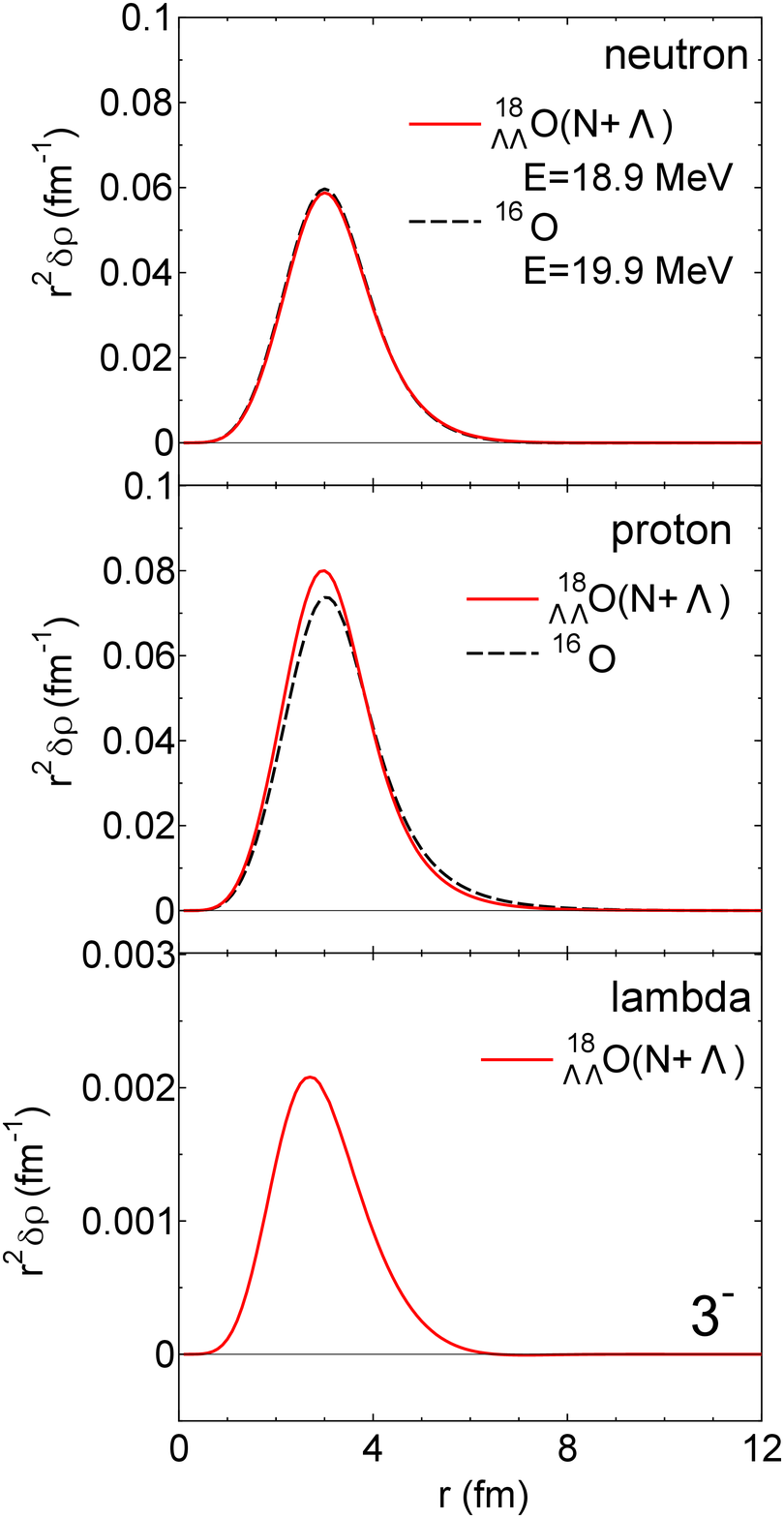}
\caption{(Color online) The transition densities for 
the giant dipole (the left panel) and the giant quadrupole (the middle panel) 
resonances and for the high-lying octupole state (the right panel) in 
$^{\,\,18}_{\Lambda\Lambda}$O (the solid lines) and $^{16}$O (the dashed lines). 
The corresponding states are denoted by the arrows in Fig. \ref{multi}.  
The top, the middle, and the bottom panels denote 
the transition densities for the neutrons, the protons and the $\Lambda$
hyperons, respectively.}
\label{drho_O16}
\end{figure*}

Figure \ref{drho_O16} shows the transition densities for 
the giant dipole and quadrupole resonances and for 
the high-lying octupole state, 
which are indicated by the arrows in Fig.\ref{multi}. 
The top, the middle and the bottom panels denote 
the transition densities for the neutrons, the protons 
and the $\Lambda$ hyperons, respectively.
Those transition densities are computed as
\begin{equation}
\begin{split}
\delta\rho_i(r)
=\sum_{ph}&(X^{(i)}_{ph}+(-1)^LY^{(i)}_{ph})\\
&\times\varphi_p(r)\varphi_h(r)\langle j_pl_p||Y_L||j_hl_h\rangle,
\end{split}
\end{equation}
where $j_i$, $l_i$ are the total and the orbital angular momenta 
for a single-particle state $i$, 
respectively, while $\varphi_i(r)$ is the radial part of the wave function,  
normalized as $\int_0^\infty \varphi_i^*(r)\varphi_i(r)\,r^2dr=1$.
The peaks of the transition densities for the neutrons and protons 
slightly move to a small distance and are shifted towards inside 
for all the multipolarities when two $\Lambda$ hyperons are added.
The amplitude of the transition density for the $\Lambda$ hyperons is about 
$1/10-1/100$ smaller than that for the protons and neutrons, so that 
the $\Lambda$ hyperons do not contribute much to these giant resonances. 
For the E2 and E3 states, the neutrons and the protons 
oscillate in phase as is expected for isoscalar motions, while they 
oscillate out of phase for the E1 state ({\it i.e.,} the isovector motion). 
In addition, the $\Lambda$ hyperons oscillate in phase with the protons 
and the neutrons for the E2 and the E3 modes, while they oscillate in phase 
with the protons for the E1 mode. 
When the Coulomb force is turned off completely, that is, when 
the single-particle levels 
for the protons are identical to those for the neutrons, 
the amplitude of 
the E1 transition density for the $\Lambda$ hyperons vanish. 

\begin{figure}
\includegraphics[width=0.80\linewidth]{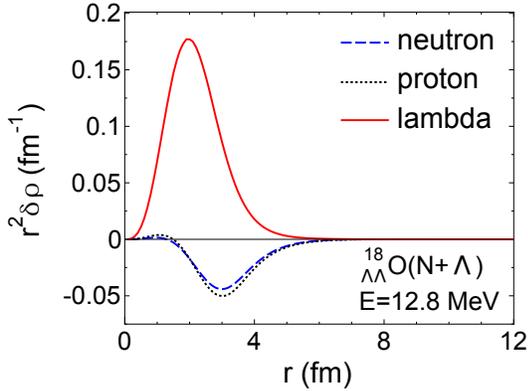}
\caption{(Color online) The transition density for the low-lying E1 state
at $E=12.8$ MeV in the $^{\,\,18}_{\Lambda\Lambda}$O hypernucleus. 
The dashed, the dotted, and the solid lines show the contributions of the 
neutrons, the protons, and the $\Lambda$ hyperons, respectively.}
\label{drho_low}
\end{figure}

The low-lying dipole state at 
$E=12.8$ MeV deserves a special attention. 
This peak appears only when the $\Lambda$ hyperons are added to 
the $^{16}$O nucleus, and a similar peak does not seen in 
other modes of excitations. 
Figure \ref{drho_low} shows the transition density for this state. 
In contrast to the giant resonance shown in Fig.\ref{drho_O16},
the amplitude of the transition density for the $\Lambda$ hyperons  
is higher than
that for the protons and the neutrons.
The strongest RPA amplitude, $\xi\equiv X^2-Y^2$, contributing to 
this peak is the $[1p\,(1s)^{-1}]$ configuration of the 
$\Lambda$ particles ($\xi=0.873$). 
The total RPA amplitudes for the neutrons and the protons 
are small ($\xi=0.050$ for the neutrons and $\xi=0.071$ for the 
protons), and these values become entirely zero 
when the $\Lambda N$ interaction is switched off. 
The neutrons and the protons oscillate in phase, 
and the $\Lambda$ particles oscillate out of phase with the nucleons. 
We can thus interpret this mode as a dipole oscillation of the $\Lambda$ 
particles against the core nucleus $^{16}$O, 
similar to the soft dipole motion of a valence 
neutron in halo nuclei \cite{Ikeda10}.

\subsection{Giant monopole resonance and incompressibility}

Giant monopole resonances, the so called ``breathing mode'', 
are intimately related to the incompressibility 
of nuclear matter 
\cite{Blaizot1995,Li2010,K09,Colo04,Khan2010,Piekarewicz2009,Blaizot1980}, 
which 
plays an important role in 
neutron stars. 
It has been shown that 
the EOS of infinite nuclear matter is 
softened 
when hyperons($\Lambda, \Xi, \Sigma$) emerge 
at high densities, and as a consequence 
the maximum mass of neutron stars 
becomes smaller \cite{Dexheimer2008,Glendenning1991,ST94}.
It is thus of interest to investigate the $\Lambda$-impurity effect of 
giant monopole resonances in finite nuclei. 

\begin{figure}
\includegraphics[width=0.70\linewidth]{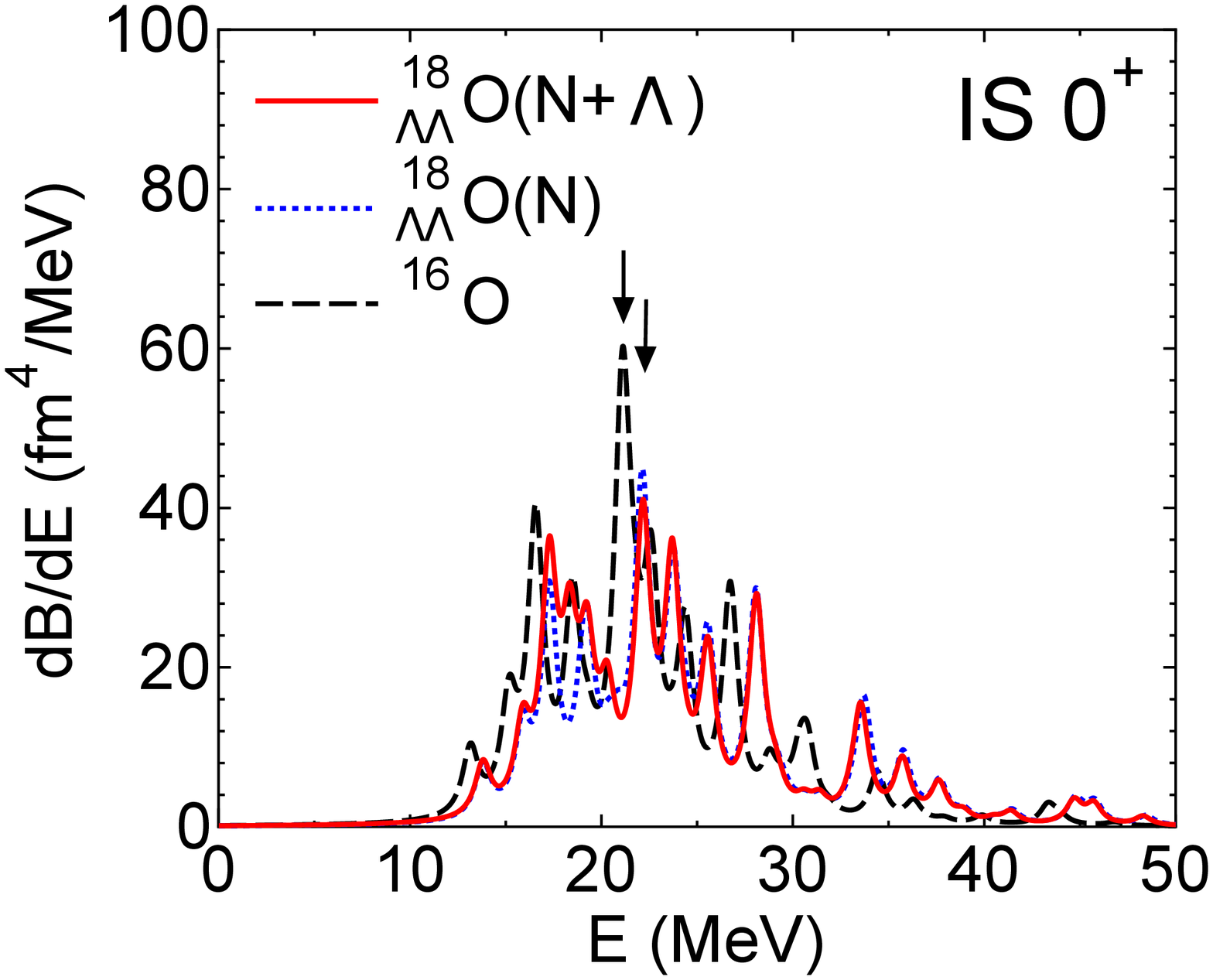}
\includegraphics[width=0.70\linewidth]{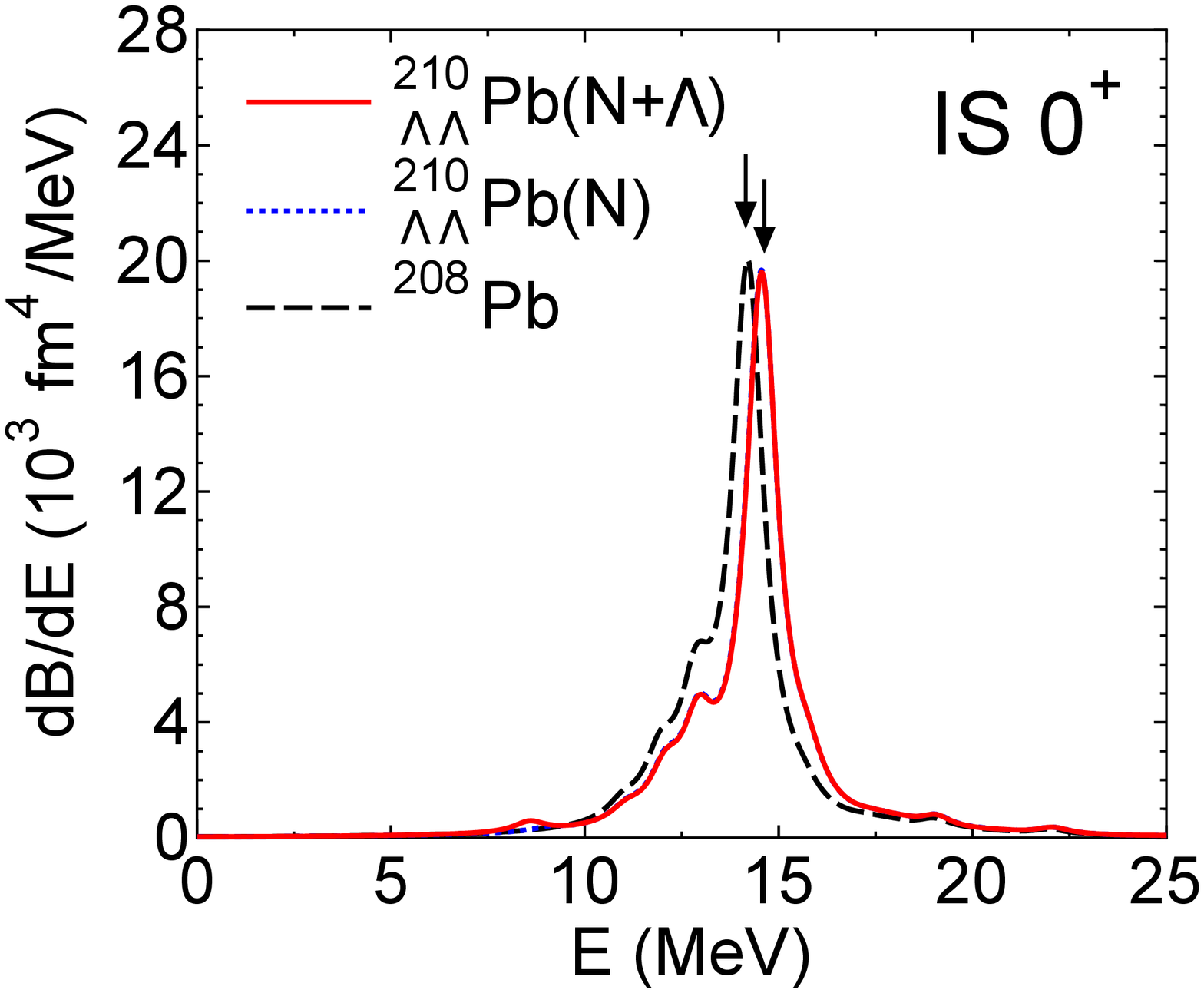}
\caption{(Color online) The strength distributions for the isoscalar 
monopole mode for 
the $^{16}$O and $^{\,\,18}_{\Lambda\Lambda}$O nuclei (the top panel) 
and for the $^{208}$Pb and $^{210}_{\Lambda\Lambda}$Pb nuclei (the bottom panel). 
The meaning of each line is the same as in Fig. \ref{multi}. }
\label{monopole}
\end{figure}

Figure \ref{monopole} shows the strength function for the isoscalar monopole 
responses of $^{16}$O and $^{\,\,18}_{\Lambda\Lambda}$O nuclei. 
The meaning of each line is the same as in Fig. \ref{multi}. 
For a comparison, the figure also shows the strength function for 
$^{208}$Pb and $^{208}_{~\Lambda\Lambda}$Pb. 
As in the other multipolarities discussed in the previous subsection, 
the strength distributions are shifted towards high energies 
when $\Lambda$ hyperons are added, and also 
the difference between the solid and the dotted lines is small, indicating 
that 
the residual $\Lambda N$ and $\Lambda\Lambda$ interactions 
play a minor role. 

\begin{figure}
\includegraphics[width=0.48\linewidth]{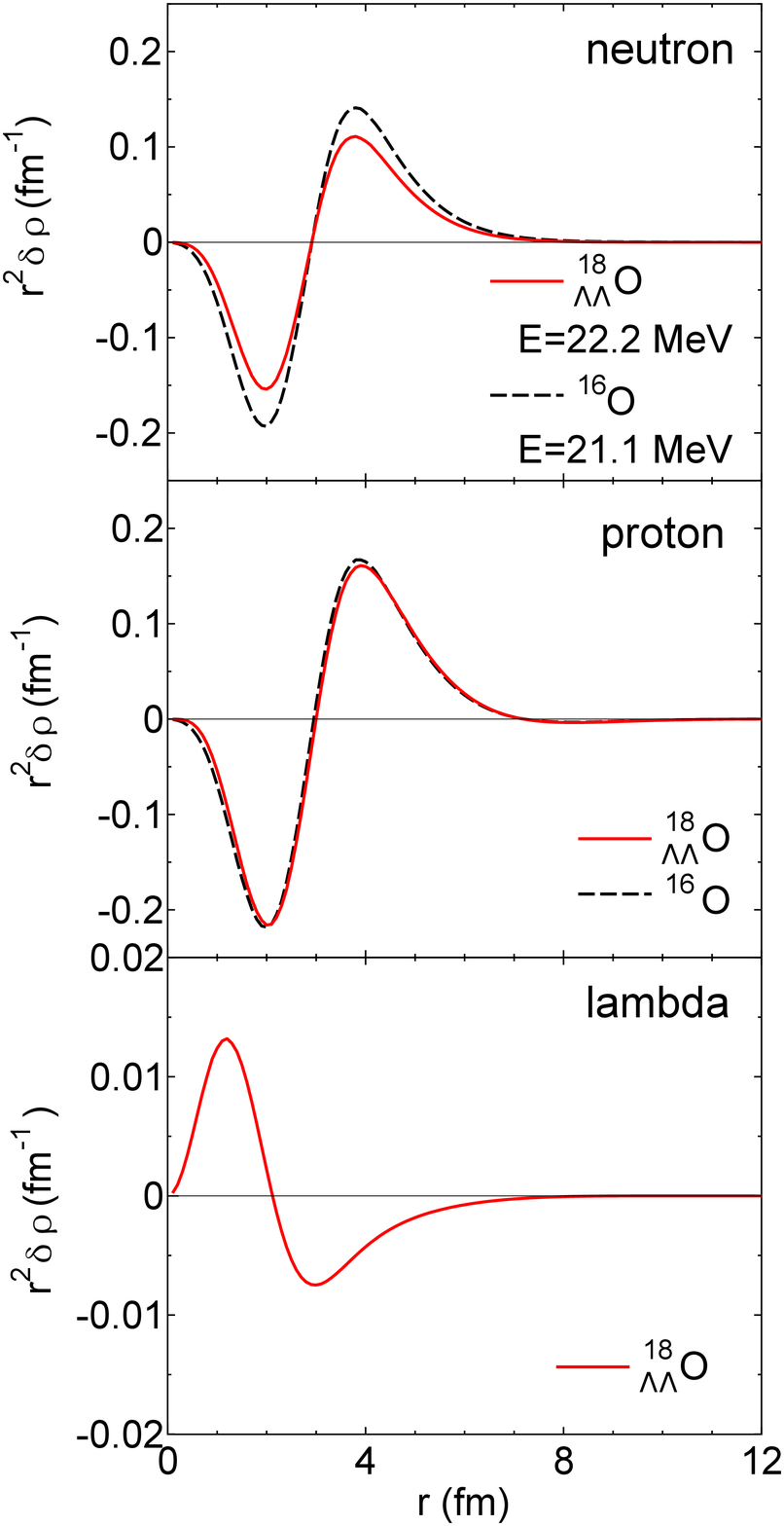}
\includegraphics[width=0.48\linewidth]{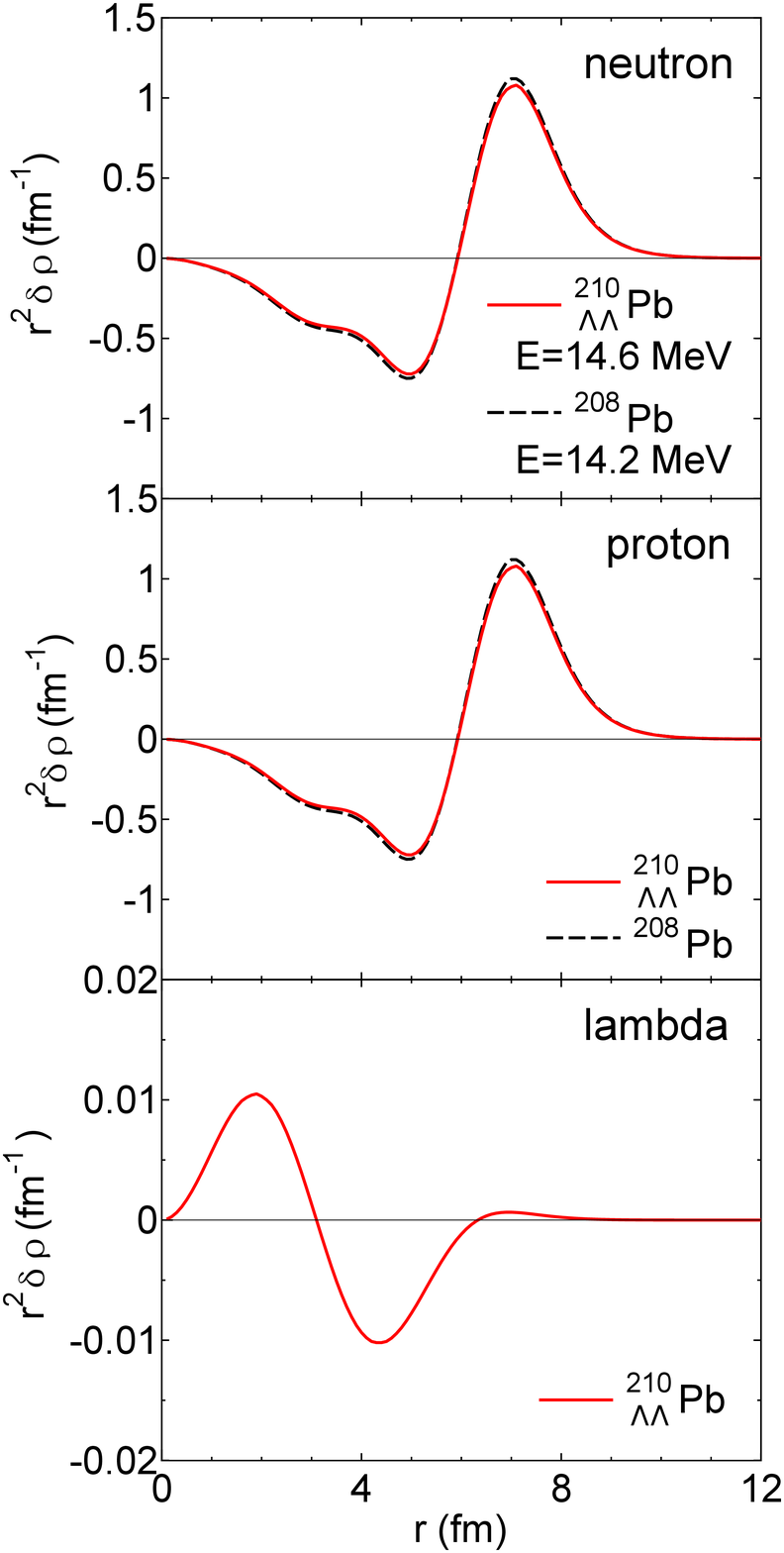}
\caption{(Color online) The transition densities for the 
isoscalar giant monopole resonances of $^{\,\,18}_{\Lambda\Lambda}$O (the left 
panels) and $^{210}_{\Lambda\Lambda}$Pb (the right panels), 
with comparisons to the transition densities for 
$^{16}$O and $^{208}$Pb. The meaning of each line is the same as in Fig. 
\ref{drho_O16}.}
\label{drho_O16GMR}
\end{figure}

Figure \ref{drho_O16GMR} shows the transition densities 
for the giant monopole resonances corresponding to the states 
indicated by the arrows in the Fig.\ref{monopole}
(that is, those states at $E=22.2$ MeV and 21.1 MeV 
for $^{\,\,18}_{\Lambda\Lambda}$O and 
$^{16}$O, respectively, and 
at 14.6 MeV and 14.2 MeV for $^{210}_{\Lambda\Lambda}$Pb and $^{208}$Pb, 
respectively).
The meaning of each line is the same as in Fig. 
\ref{drho_O16}. 
For the oxygen nuclei, when $\Lambda$ hyperons are added, 
the amplitude of the transition density for the neutrons 
decreases by about $20\%$ while that for the protons 
remains almost the same.  
The amplitude of the $\Lambda$ transition density is about 10 times 
smaller than that of the nucleons. 
It is interesting to notice that the $\Lambda$ hyperons oscillate 
out of phase with the nucleons. 
These features are qualitatively the same for the lead nuclei as well, 
although the changes in the transition densities are much smaller compared 
to the oxygen isotopes. 

According to Blaizot, 
the effective incompressibility 
modulus $K_A$ for finite nuclei 
without $\Lambda$ hyperons is defined as \cite{Blaizot1980}, 
\begin{equation}
K_A=\frac{m_N}{\hbar^2}\mathcal{E}^2\langle r^2 \rangle,
\label{blaizot1}
\end{equation}
where $\sqrt{\langle r^2 \rangle}$ is the root mean square radius, 
and $\mathcal{E}^2=m_{1}/m_{-1}$ (see Eq. (\ref{ewsr}) for the definition of 
the $k$-th energy-weighted sum-rule, $m_k$). 
When the $\Lambda$ hyperons are present, 
this formula is modified as, 
\begin{equation}
K_A=\frac{m_N}{\hbar^2}\mathcal{E}^2\langle r^2 \rangle
\left(
\frac{N+Z}{A}\frac{\langle r^2 \rangle_{\rm n+p}}
{\langle r^2 \rangle} 
+
\frac{m_NN_\Lambda}{m_\Lambda A}\frac{\langle r^2 \rangle_\Lambda}
{\langle r^2 \rangle} 
\right)^{-1},
\label{blaizot2}
\end{equation}
where we have used 
Eqs. (3.45) and (3.47) in Ref.\cite{Blaizot1980}
and 
the energy-weighted sum-rule for 
the isoscalar monopole transition, 
\begin{equation}
m_1(L=0)=
\frac{2\hbar^2}{m_N}(N+Z)\langle r^2 \rangle_{\rm{n+p}}
+
\frac{2\hbar^2}{m_\Lambda}N_\Lambda\langle r^2 \rangle_{\Lambda}.
\label{ewsr1}
\end{equation}
In Eq. (\ref{blaizot2}), 
$\sqrt{\langle r^2 \rangle}_{\rm{n+p}}$ is 
the root mean square radius of the core nucleus.
Notice that Eq. \eqref{blaizot2} is reduced to 
Eq.\eqref{blaizot1} 
when $N_\Lambda=0$.
In Table \ref{centroid0}, we list the centroid energy $\overline{E}_{0^+}$,
$\mathcal{E}$ for the isoscalar monopole modes, 
the root-mean-square radii, $\sqrt{\langle r^2 \rangle}$ 
and $\sqrt{\langle r^2 \rangle_{\rm n+p}}$, and 
the effective incompressibility, $K_A$, calculated according to 
Eqs. (\ref{blaizot1}) and (\ref{blaizot2}). 
When $\Lambda$ hyperons are added, the centroid energies increase 
by $1.9$ MeV for $^{16}$O and $0.4$ MeV for $^{208}$Pb, and 
the rms radii for the core nucleus, $\sqrt{\langle r^2\rangle_{\rm{n+p}}}$, 
decrease by $0.04$ fm for $^{16}$O 
and $0.01$ fm for $^{208}$Pb. 
As we have shown, the increase of the centroid energies is 
mainly due to the change of single-particle levels, and 
the residual $\Lambda N$ and $\Lambda\Lambda$
interactions give only a minor effect. 
The decrease of the rms radii 
is attributed to the attractive $\Lambda N$ interaction, that is, the 
shrinkage effect of $\Lambda$ hyperons. 
The effective incompressibility, $K_A$, increases for both 
the nuclei studied when $\Lambda$ hyperons are added.

\begin{table}
\caption{
Properties of the isoscalar monopole responses obtained with the 
Skyrme HF+RPA method. 
$\overline{E}_{0^+}$ is the centroid energy, and $\mathcal{E}$ 
is defined as 
$\mathcal{E}=\sqrt{m_1/m_{-1}}$. 
$\sqrt{\langle r^2 \rangle_{\rm{n+p}}}$ 
and $\sqrt{\langle r^2 \rangle}$ are 
the root mean square radii for the core nuclei and that for 
the total densities, respectively. 
$K_A$ is the effective nuclear incompressibility defined by 
Eqs. (\ref{blaizot1}) and (\ref{blaizot2}).}
\begin{tabular}{ccccc|c}
\hline\hline 
 & $\overline{E}_{0^+}$(MeV) &$\mathcal{E}$(MeV) 
 & $\sqrt{\langle r^2 \rangle_{\rm{n+p}}}$ & $\sqrt{\langle r^2 \rangle}$(fm)
 & $K_A$(MeV)\\
\hline
$^{16}$O                    & $22.4$ & $21.7$ & $2.68$ & $2.68$ & $81.6$ \\
$^{\,\,18}_{\Lambda\Lambda}$O & $24.3$ & $23.5$ & $2.64$ & $2.58$ & $90.0$ \\
\hline
$^{208}$Pb                & $14.1$ & $14.0$ & $5.56$ & $5.56$ & $146$ \\
$^{210}_{\Lambda\Lambda}$Pb & $14.5$ & $14.4$ & $5.55$ & $5.53$ & $153$ \\
\hline\hline 
\end{tabular}
\label{centroid0}
\end{table}
\begin{figure}
\includegraphics[scale=0.4]{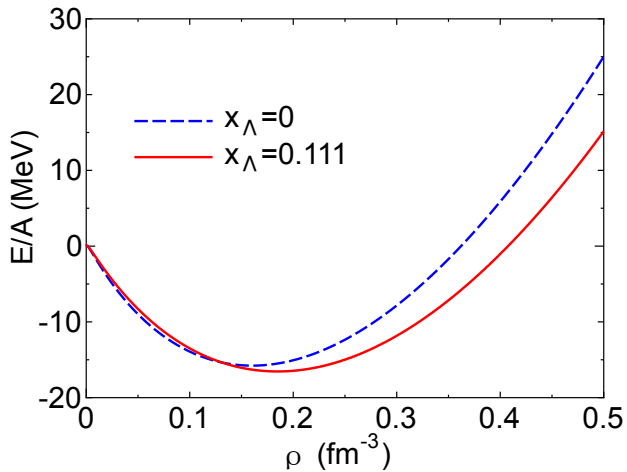}
\caption{(Color online) The 
binding energy per particle in infinite nuclear matter for the fraction of 
$\Lambda$ particle of $x_\Lambda$=0 (the dashed line) and 
$x_\Lambda$=0.11 (the solid line). The neutron and the proton densities are 
set to be equal, that is, $\rho_n=\rho_p=(1-x_\Lambda)\rho/2$. }
\end{figure}

The increase of the effective incompressibility should reflect the properties 
of infinite nuclear matter. 
In order to assess this, Fig. 7 shows the binding energy 
per particle in infinite nuclear matter, $E/A$, 
given as 
\begin{eqnarray} 
\frac{E}{A}&=&\frac{H}{\rho}
=\frac{H_N}{\rho}+\frac{1}{\rho}\left[\frac{\hbar^2}{2m_\Lambda}\tau_\Lambda
+t_0^\Lambda\left(1+\frac{x_0^\Lambda}{2}\right)\rho_N\rho_\Lambda \right.\nonumber \\
&&+\frac{3}{8}t_3^\Lambda\rho_\Lambda\rho_N^2
+\frac{1}{4}(t_1^\Lambda+t_2^\Lambda)(\tau_\Lambda\rho_N+\tau_N\rho_\Lambda) \nonumber \\
&&+\frac{\lambda_0}{4}\rho_\Lambda^2+\frac{1}{8}(\lambda_1+3\lambda_2)\rho_\Lambda\tau_\Lambda \nonumber \\
&&\left.+\frac{\lambda_3}{4}\rho_\Lambda^2\rho_N^\alpha\right],
\end{eqnarray}
where $H_N$ is the nucleon part of the 
energy density evaluated in infinite matter. 
$\rho$ is the total density, while $\rho_\Lambda\equiv x_\Lambda \rho$
and $\rho_N\equiv (1-x_\Lambda) \rho$ are the densities of $\Lambda$ particles 
and nucleons, respectively. 
The kinetic energy densities $\tau_N$ and $\tau_\Lambda$ are evaluated as 
$\tau_N=3\rho_N k_{FN}^2/5$ and $\tau_\Lambda=3\rho_\Lambda k_{F\Lambda}^2/5$, respectively, where 
the Fermi momenta are given by $k_{FN}=(3\pi^2\rho_N/2)^{1/3}$ and  
$k_{F\Lambda}=(3\pi^2\rho_\Lambda)^{1/3}$.
The dashed and the solid lines in Fig. 7 show the results 
with the $\Lambda$ fraction of $x_\Lambda$=0 and $x_\Lambda$=2/18, respectively. 
Here, we have assumed that the neutron and the proton densities are 
the same, 
$\rho_n=\rho_p=\rho_N/2$. 
One can see that the addition of $\Lambda$ particles shifts the equilibrium 
density $\rho_0$ towards a high density, that is, $\rho_0=0.161$ fm$^{-3}$ for 
$x_\Lambda$=0 and $\rho_0=0.185$ fm$^{-3}$ for $x_\Lambda$=2/18. 
The incompressibility for infinite nuclear matter, 
\begin{equation}
K_\infty=9\rho_0^2\,\left(\frac{d^2(E/A)}{d\rho^2}\right)_{\rho=\rho_0},
\end{equation}
is $K_\infty=217$ MeV for $x_\Lambda$=0 and $K_\infty=239$ MeV 
for $x_\Lambda=2/18$, agreeing with the increase of 
effective incompressibility of the finite hypernuclei.

Notice that this observation does not contradict to the fact that 
hyperons soften the equation of state for infinite nuclear matter 
relevant to neutron stars. 
That is, in neutron star calculations, the emergence of hyperons 
takes place at high densities and nucleons are the only constituents 
at the normal density, when the beta stability condition is imposed. 
In contrast, Fig. 7 shows the effect of hyperons on the 
incompressibility defined at the equilibrium density. 
Even though the beta stability condition does not hold 
there if the $\Lambda$ fraction is finite, 
this EOS is more relevant to giant monopole resonances of finite hypernuclei.

\section{SUMMARY}

We have extended the Skyrme-HF plus RPA schemes 
to calculations for vibrational excitations of double-$\Lambda$ hypernuclei.
We have applied it to the electric dipole (E1), the quadrupole (E2), and 
the octupole (E3) modes of excitations in the $^{\,\,18}_{\Lambda\Lambda}$O 
hypernucleus. 
We have shown that the strength distributions shift towards high energies 
for all the modes when the $\Lambda$ hyperons are added to $^{16}$O. 
This is the case both for the low-lying quadrupole and octupole states 
and for giant resonances. 
At the same time, the electromagnetic transition probabilities also decrease. 
We have argued that these features are mainly 
caused by the change in the single-particle energies, whereas 
the residual $\Lambda N$ and $\Lambda\Lambda$ interactions play a minor role. 
The calculated transition densities show that 
the peak of the transition densities are shifted towards inside
and the height of the peaks slightly changes due to the impurity effect 
of $\Lambda$ hyperons.

For the E1 strength, we have found a new peak 
at low-lying energy, that is absent 
in the E1 response of the core nucleus. 
From the analysis of the transition density, 
we have shown that this state corresponds to an oscillation of 
the $\Lambda$ particles against the core nucleus. 

We have also discussed the 
$\Lambda$-impurity effect on the isoscalar monopole 
vibration 
of $^{\,\,18}_{\Lambda\Lambda}$O and 
$^{210}_{\Lambda\Lambda}$Pb, 
and the incompressibility of infinite nuclear matter in the 
presence of $\Lambda$ hyperons. 
When the $\Lambda$ hyperons are added, the strength 
distributions are shifted to higher energies and thus 
the centroid energies increase, 
similarly to the other multipole transitions. 
We have shown that the transition density for the $\Lambda$ particles 
behave rather differently from the transition densities for the 
neutrons and the protons. 
The increase of the centroid energy for the giant monopole resonance 
implies that $\Lambda$ particles increase the 
nuclear incompressibility, when hyperons were emerged at 
the equilibrium density. 

In this paper, we have studied several collective vibrational motions, taking 
the $^{\,\,18}_{\Lambda\Lambda}$O hypernucleus as examples. 
It would be an interesting future work to 
study systematically the $\Lambda$-impurity effect 
on the collective excitations of other double-$\Lambda$ hypernuclei. 
In particular, the low-lying dipole mode, originated from a dipole 
oscillation of the $\Lambda$ particles against the core nucleus, would 
be interesting to study. For this purpose, we would have to extend our 
formalism by including the pairing correlations with the quasi-particle 
RPA (QRPA). 
Another interesting extension is to study the collective excitations of 
single-$\Lambda$ hypernuclei, although 
the broken time-reversal symmetry will have to be taken into account 
correctly there.

\begin{acknowledgments}
We thank G. Colo and H. Sagawa for discussions. 
This work was supported by the Japanese
Ministry of Education, Culture, Sports, Science and Technology
by Grant-in-Aid for Scientific Research under
the program number (C) 22540262.
\end{acknowledgments}

\appendix 

\section{Energy densities and mean-field potentials for hypernuclei}
\renewcommand{\theequation}{A.\arabic{equation}}
\setcounter{equation}{0}
In this Appendix A, we summarize the explicit formulae for the 
energy densities $H_{N\Lambda}$ and $H_{\Lambda\Lambda}$ 
in Eq.\eqref{energy3} and the mean-field potentials 
in Eq.\eqref{shfeq1}. 

The energy density $H_{N\Lambda}$, due to the Skyrme-type 
$\Lambda N$ interaction given by Eq.\eqref{LN}, reads \cite{Rayet},
\begin{equation}
\begin{split}
&H_{N\Lambda}(\vec{r})=
\frac{\hbar^2}{2m_\Lambda}\tau_\Lambda+t_0^\Lambda
\left(1+\frac{1}{2}x_0^\Lambda\right)
\rho_N\rho_\Lambda\\
&+\frac{1}{4}(t_1^\Lambda+t_2^\Lambda)(\tau_\Lambda\rho_N+\tau_N\rho_\Lambda)
+\frac{1}{8}(3t_1^\Lambda-t_2^\Lambda)
(\vec{\nabla}\rho_N\cdot\vec{\nabla}\rho_\Lambda)\\
&+\frac{1}{2}W_0^\Lambda(\vec{\nabla}\rho_N\cdot\vec{J}_\Lambda
+\vec{\nabla}\rho_\Lambda\cdot\vec{J}_N)
+\frac{1}{4}t_3^\Lambda\rho_\Lambda(\rho_N^2+2\rho_n\rho_p), 
\end{split}
\label{eln}
\end{equation}
while the $\Lambda\Lambda$ part, $H_{\Lambda\Lambda}$, 
originated from the $\Lambda\Lambda$ interaction given by 
Eq.\eqref{LL} reads\cite{Lanskoy1998}, 
\begin{equation}
\begin{split}
&H_{\Lambda\Lambda}(\vec{r})=
\frac{1}{4}\lambda_0\rho_\Lambda^2
+\frac{1}{8}(\lambda_1+3\lambda_2)\rho_\Lambda\tau_\Lambda\\
&+\frac{3}{32}(\lambda_2-\lambda_1)\rho_\Lambda\vec{\nabla}^2\rho_\Lambda
+\frac{1}{4}\lambda_3\rho_\Lambda^2\rho_N^\alpha.
\end{split}
\label{ell}
\end{equation}
Here, $\rho_b=\rho_b(\vec{r})$, $\tau_b=\tau_b(\vec{r})$, 
$\vec{J}_b=\vec{J}_b(r)$ are the number, the kinetic energy and the 
spin densities, respectively ($b=p,n$, or $\Lambda$).
The indices $N$, $p$, $n$ and $\Lambda$ are the nucleon, the proton, the 
neutron
and the $\Lambda$ hyperon, respectively.

After taking variation of the energy in Eq.\eqref{energy} with respect to 
the densities, we obtain the Skyrme-Hartree-Fock 
equation given by Eq.\eqref{shfeq1}.
The mean-field potentials in Eq.\eqref{shfeq1} are given by 
\begin{equation}
\begin{split}
&U_{q\Lambda}(\vec{r})
=t_0^\Lambda\left(1+\frac{1}{2}x_0^\Lambda\right)\rho_\Lambda
+\frac{1}{4}\left(t_1^\Lambda+t_2^\Lambda\right)\tau_\Lambda\\
&-\frac{1}{8}\left(3t_1^\Lambda-t_2^\Lambda\right)\vec{\nabla}^2\rho_\Lambda
-\frac{1}{2}W_0^\Lambda\vec{\nabla}\cdot\vec{J}_\Lambda\\
&+\frac{1}{2}W_0^\Lambda\vec{\nabla}\rho_\Lambda\cdot
(-i\vec{\nabla}\times\vec{\sigma})
+\frac{1}{2}t_3^\Lambda\rho_\Lambda(2\rho_N-\rho_q),
\end{split}
\end{equation}
\begin{equation}
\begin{split}
&U_{\Lambda N}(\vec{r})
=t_0^\Lambda\left(1+\frac{1}{2}x_0^\Lambda\right)\rho_N
+\frac{1}{4}\left(t_1^\Lambda+t_2^\Lambda\right)\tau_N\\
&-\frac{1}{8}\left(3t_1^\Lambda-t_2^\Lambda\right)\vec{\nabla}^2\rho_N
-\frac{1}{2}W_0^\Lambda\vec{\nabla}\cdot\vec{J}_N\\
&+\frac{1}{2}W_0^\Lambda\vec{\nabla}\rho_N\cdot
(-i\vec{\nabla}\times\vec{\sigma})
+\frac{1}{4}t_3^\Lambda(\rho_N^2+2\rho_n\rho_p),
\end{split}
\end{equation}
and
\begin{equation}
\begin{split}
U_{\Lambda\Lambda}(\vec{r})
&=\frac{1}{2}\lambda_0\rho_\Lambda
+\frac{1}{8}(\lambda_1+3\lambda_2)\tau_\Lambda\\
&+\frac{3}{16}(\lambda_2-\lambda_1)\vec{\nabla}^2\rho_\Lambda
+\frac{1}{2}\lambda_3\rho_\Lambda\rho_N^\alpha. 
\end{split}
\end{equation}
Note that the index $q$ refers only to the proton and the neutron.
The effective mass for the nucleons and the $\Lambda$ hyperons 
in Eq. \eqref{shfeq1} 
are given by 
\begin{equation}
\frac{\hbar^2}{2m_q^*}=\frac{\hbar^2}{2m_N}
+\frac{1}{4}(t_1^\Lambda+t_2^\Lambda)\rho_\Lambda(\vec{r}),
\label{effect_m1}
\end{equation}
and
\begin{equation}
\frac{\hbar^2}{2m_{\Lambda}^*}=\frac{\hbar^2}{2m_\Lambda}
+\frac{1}{4}(t_1^\Lambda+t_2^\Lambda)\rho_N(\vec{r})
+\frac{1}{8}(\lambda_1+3\lambda_3)\rho_\Lambda(\vec{r}),
\label{effect_m2}
\end{equation}
respectively. 

\section{$\Lambda N$ and $\Lambda\Lambda$ residual interactions}
\renewcommand{\theequation}{B.\arabic{equation}}
\setcounter{equation}{0}
\label{appb}

The matrix elements for a particle-hole residual interaction $v_{\rm res}$ 
are given as\cite{RingSchuck,Rowe}
\begin{equation}
\begin{split}
v_{ph'hp'}&=\langle p(h)^{-1}LK|v_{\rm{res}}|p'(h')^{-1}LK\rangle,\\
v_{pp'hh'}&=\langle p(h)^{-1}LK,p'(h')^{-1}L\bar{K}|v_{\rm{res}}|\rangle,
\end{split}
\end{equation}
where $L$ is the multipolarity for the particle-hole excitations and $K$ is 
its $z$-component. 
For hypernuclei, 
the residual interaction 
can be separated into two parts, 
$v_{\rm{res}}=v_{\rm{res}}^{b_1b_2}(N)+v_{\rm{res}}^{b_1b_2}(\Lambda)$, 
where the indices $b_1$ and $b_2$ denote $p$, $n$ or $\Lambda$. 
The interaction $v_{\rm{res}}^{b_1b_2}(N)$ is due to 
the $NN$ residual interaction, 
whose explicit form can be found in {\it e.g.}
Refs.\cite{Liu1991,Hamamoto1999,Terasaki2005}. 
$v_{\rm{res}}^{b_1b_2}(\Lambda)$ is the additional term
due to the $\Lambda N$ and the $\Lambda\Lambda$ residual interactions. 
These are given in the form of, 
\begin{equation}
\begin{split}
&v_{\rm{res}}^{b_1b_2}(\Lambda)=\delta(\vec{r}_{b1}-\vec{r}_{b2})\Big(a_{b_1b_2}\\
&+b_{b_1b_2}\Big[\vec{\nabla}_1^2 +\vec{\nabla}_2^2
                           +\vec{\nabla}_1'^2+\vec{\nabla}_2'^2
-(\vec{\nabla}_1-\vec{\nabla}_1')(\vec{\nabla}_2-\vec{\nabla}_2')\Big]\\
&+c_{b_1b_2}(\vec{\nabla}_1+\vec{\nabla}_1')(\vec{\nabla}_2+\vec{\nabla}_2')\Big), 
\end{split}
\label{ph-int}
\end{equation}
where $a_{b_1b_2}$, $b_{b_1b_2}$ and $c_{b_1b_2}$ are given by 
\begin{equation}
\begin{split}
a_{qq'}=&\frac{t_3^\Lambda}{4}\rho_{\Lambda}
\Big(3-\vec{\sigma}\cdot\vec{\sigma}'-\vec{\tau}\cdot\vec{\tau}'
-\vec{\sigma}\cdot\vec{\sigma}'\vec{\tau}\cdot\vec{\tau}'\Big)\\
&+\frac{\lambda_3}{4}\alpha(\alpha-1)\rho_N^{\alpha-2}\rho_\Lambda^2\\
b_{qq'}&=c_{qq'}=0,
\end{split}
\end{equation}
\begin{equation}
\begin{split}
a_{\Lambda q}&=t_0^\Lambda\left(1+\frac{x_0^\Lambda}{2}\right)
+\frac{t_0^\Lambda x_0^\Lambda}{2}\vec{\sigma}_1\cdot{\vec{\sigma}_2}\\
&+t_3^\Lambda\left(\rho_N-\frac{\rho_q}{2}\right)
+\frac{\lambda_3}{2}\alpha\rho_N^{\alpha-1}\rho_\Lambda\\
b_{\Lambda q}&=-\frac{1}{8}(t_1^\Lambda+t_2^\Lambda),\quad
c_{\Lambda q}=\frac{1}{8}(t_1^\Lambda-3t_2^\Lambda).\\
\end{split}
\end{equation}
for ($q,q'=p$ or $n$), and 
\begin{equation}
\begin{split}
a_{\Lambda\Lambda}&
=\frac{1}{2}\lambda_0(1-\vec{\sigma}_1\cdot\vec{\sigma}_2)
+\frac{1}{2}\lambda_3\rho_N^\alpha(1-\vec{\sigma}_1\cdot\vec{\sigma}_2)\\
b_{\Lambda\Lambda}&=-\frac{1}{16}\Big(
\lambda_1(1-\vec{\sigma}_1\cdot\vec{\sigma}_2)
+\lambda_2(3+\vec{\sigma}_1\cdot\vec{\sigma}_2)\Big)\\
c_{\Lambda\Lambda}&=\frac{1}{16}\Big(
\lambda_1(1-\vec{\sigma}_1\cdot\vec{\sigma}_2)
-3\lambda_2(3+\vec{\sigma}_1\cdot\vec{\sigma}_2)\Big), 
\end{split}
\end{equation}
for the $\Lambda\Lambda$ terms.

\end{document}